\documentclass[conference]{IEEEtran}
\usepackage{textcomp}
\usepackage{graphicx}
\usepackage{xcolor}

\usepackage{caption}
\usepackage[caption=false,font=footnotesize,labelfo
nt=sf,textfont=sf]{subfig}

\usepackage{cite}
\usepackage[tbtags]{amsmath}
\usepackage{amssymb,amsfonts}
\usepackage{algorithmic}
\usepackage{textcomp}
\usepackage{multirow}
\usepackage[normalem]{ulem}

\usepackage{cite}
\usepackage{CJK}

\usepackage{cleveref}
\usepackage{booktabs}
\usepackage{steinmetz}
\usepackage{url}
\usepackage{flushend}

\usepackage[T1]{fontenc} 
\usepackage{amsmath}
\usepackage[cmintegrals]{newtxmath}
\usepackage{bm}
\usepackage{array}

\usepackage{url}

\ifCLASSINFOpdf
\else
\fi

\providecommand{\hypersetup}[1]{\relax}

\hypersetup{pdftitle={Bare Demo of IEEE\_lsens.cls for IEEE Sensors Letters},
pdfsubject={Typesetting},
pdfauthor={Michael D. Shell},
pdfkeywords={Class, IEEE, IEEE\_lsens, IEEE Sensors Letters, LaTeX, Typesetting, TeX}}

\begin{document}
\title{Thin Film Reconfigurable Intelligent Surface for Harmonic Beam Steering}
\author{\IEEEauthorblockN{Boxuan Xie\IEEEauthorrefmark{1}, Aleksandr D. Kuznetsov\IEEEauthorrefmark{2}, Lauri Mela\IEEEauthorrefmark{1},  Jari Lietzén\IEEEauthorrefmark{1}, Kalle Ruttik\IEEEauthorrefmark{1}, Alp Karako\c{c}\IEEEauthorrefmark{1},
and Riku Jäntti\IEEEauthorrefmark{1}}
\IEEEauthorblockA{\IEEEauthorrefmark{1}Department of Information and Communications Engineering,
Aalto University, Espoo, 02150, Finland\\
\IEEEauthorrefmark{2}Department of Electronics and Nanoengineering,
Aalto University, Espoo, 02150, Finland\\
}
}
\IEEEtitleabstractindextext{%
\begin{abstract}
This letter explores an implementation of a novel thin film $1\times4$ reconfigurable intelligent surface (RIS) designed for future communication and sensing scenarios. Utilizing cost-effective inkjet printing methods and additive manufacturing, our approach significantly simplifies the RIS construction process and reduces production costs. The RIS, fabricated on a flexible and lightweight polyethylene terephthalate (PET) substrate, integrates antennas, switching circuitry, and a microcontroller unit (MCU), without a ground shield. This setup enables individual and simultaneous control of each RIS element, manipulating the captured carrier signal by reflecting and refracting its dominant harmonics. Beams of the harmonics can be steered to multiple desired directions at both front and back sides of the surface. Measurement results of the beam steering show that the RIS has the potential to enable RIS-aided communication and sensing applications.
\end{abstract}
}
\IEEEpubid{
\\1949-307X \copyright\ 2023 IEEE. Personal use is permitted, but republication/redistribution requires IEEE permission.\\
See \url{http://www.ieee.org/publications\_standards/publications/rights/index.html} for more information.}
\maketitle
\section{Introduction}
\IEEEPARstart{I}{n} the landscape of 6G wireless communication systems, reconfigurable intelligent surface (RIS) is emerging as a crucial technology that can enhance the control and manipulation of electromagnetic waves~\cite{marco2020tutorial}.
RIS can adjust the phase, amplitude, and polarization of incoming waves, which not only improves signal quality and coverage but also boosts the sensing capabilities in complex environments.
The application of RIS in sensing systems is particularly promising~\cite{zhang2022sensing}.
Fig.~\ref{fig:ris_concept_sensing} depicts two common RIS-aided sensing scenarios.
For device-based sensing, RIS can extend the functional range and accuracy of sensors embedded within communication devices, leading to better data capture and interaction with the environment. 
In device-free sensing, RIS enables the detection and analysis of signal disruptions caused by objects or events in the environment, which facilitates monitoring without direct sensor placements.

Recent advancements in RIS prototyping have led to a significant expansion beyond conventional reflective features. 
For instance, research on holographic surfaces~\cite{huang2020holographic} introduces capabilities that include relay-type and transmitter-type operations, alongside both discrete and continuous implementations.
To extend the signal coverage of reflection-based intelligent reflecting surface (IRS), the intelligent omni-surface (IOS)~\cite{zhang2022ios} and the simultaneous transmitting and reflecting RIS (STAR-RIS)~\cite{mu2022star} were proposed to enable both signal reflection and refraction of RISs.
However, conventional RIS elements have been constructed using multiple layers of conductive and dielectric materials stacked together. This approach, while effective, typically necessitates complex manufacturing processes that can elevate the cost of devices.
Furthermore, conventional RIS requires complicated and power-hungry control circuits, such as FPGA-based controllers.
To address these challenges, recent advancements in additive manufacturing of backscatter devices with cost-effective inkjet printing methods are being leveraged~\cite{song2022survey,costa2021review}.

This letter presents a low-cost and low-complexity thin film $1\times4$ RIS with both reflective and refractive features working at 2.4~GHz industrial, scientific, and medical (ISM) band, shown in Fig.~\ref{fig:ris_proto}. The RIS is fabricated using inkjet printing technique and additive manufacturing methodology with commercial off-the-shelf materials.
The RIS components, such as antennas, switching circuitry, and a low-power microcontroller unit (MCU) are integrated on a flexible and lightweight polyethylene terephthalate (PET) substrate, without a ground shield. 
The MCU can separately and simultaneously control the load modulation of each RIS element with the MCU-generated phased baseband signals by switching operation.
The load modulation thus leads to the formation of harmonics into the captured radio frequency (RF) carrier signal.
The manufactured RIS can effectively control the reflecting and refracting of dominant harmonics by steering beams to desired directions at both front and back sides of the surface.
\section{System Model}
\begin{figure}
\centering
\subfloat[]
{\includegraphics[width=0.48\columnwidth]{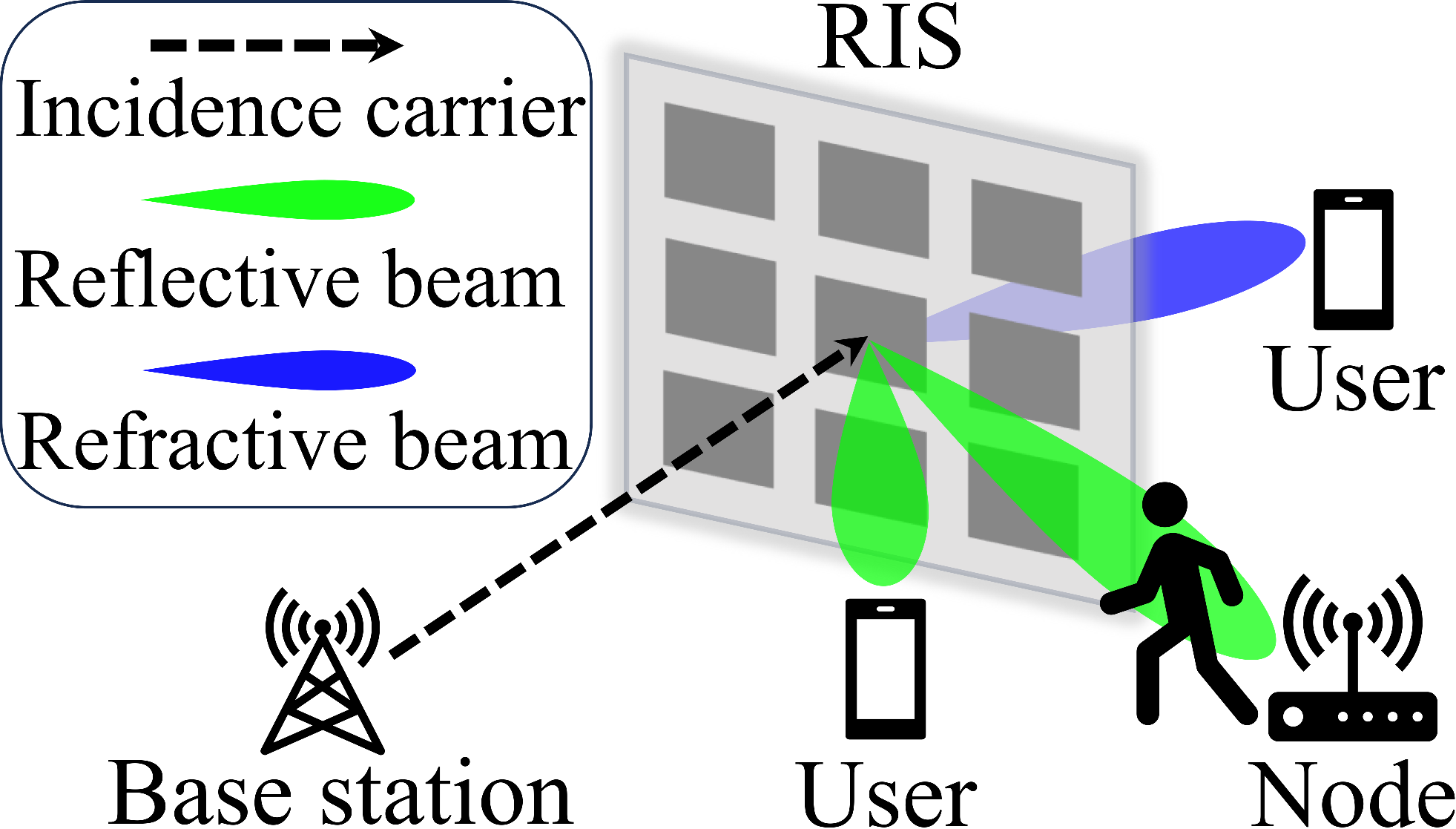}
    \label{fig:ris_concept_sensing}}
\hfill
\centering
\subfloat[]
{\includegraphics[width=0.48\columnwidth]{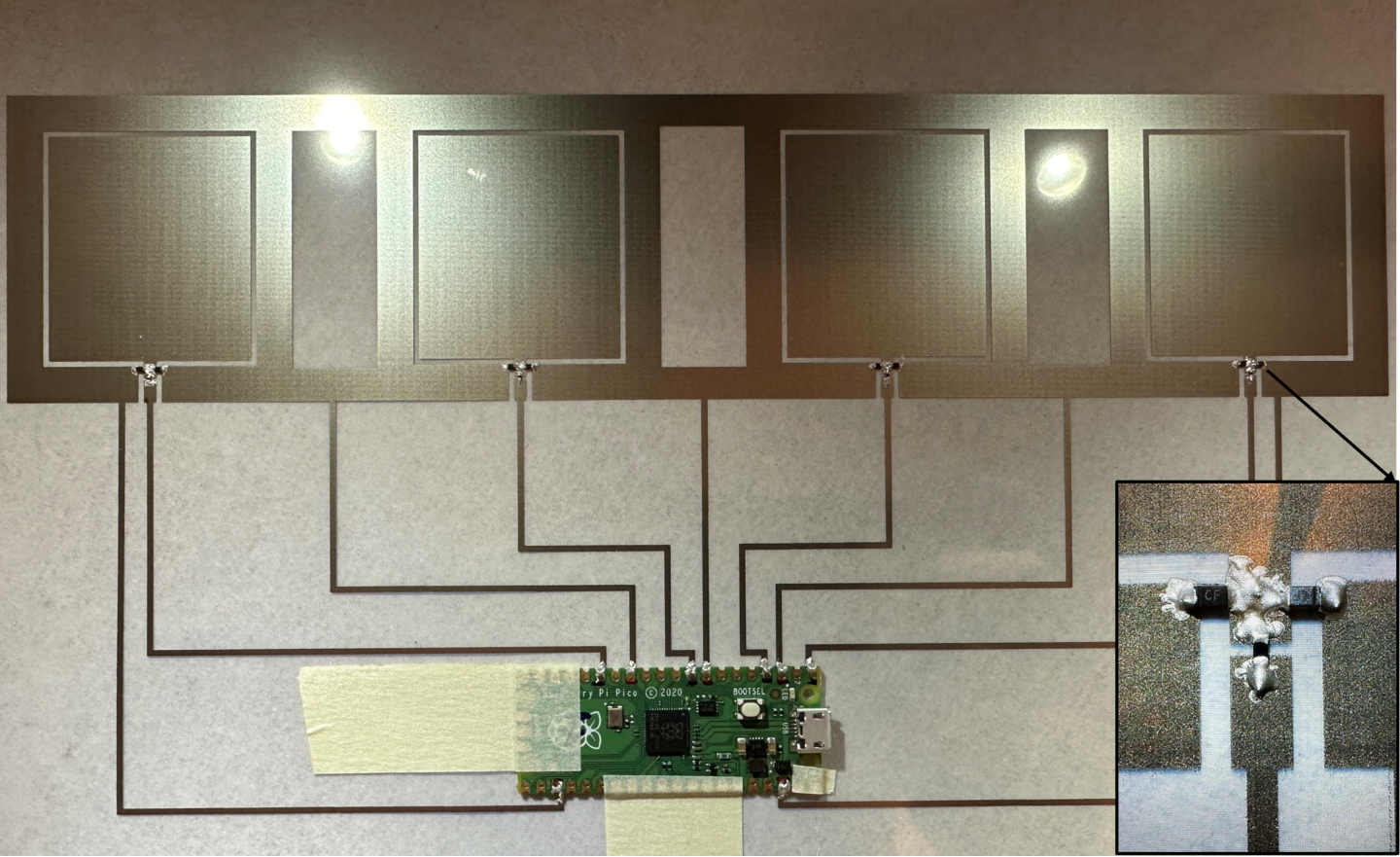}
    \label{fig:ris_proto}}
\hfill
\centering
\subfloat[]
    {\includegraphics[width=0.98\columnwidth]{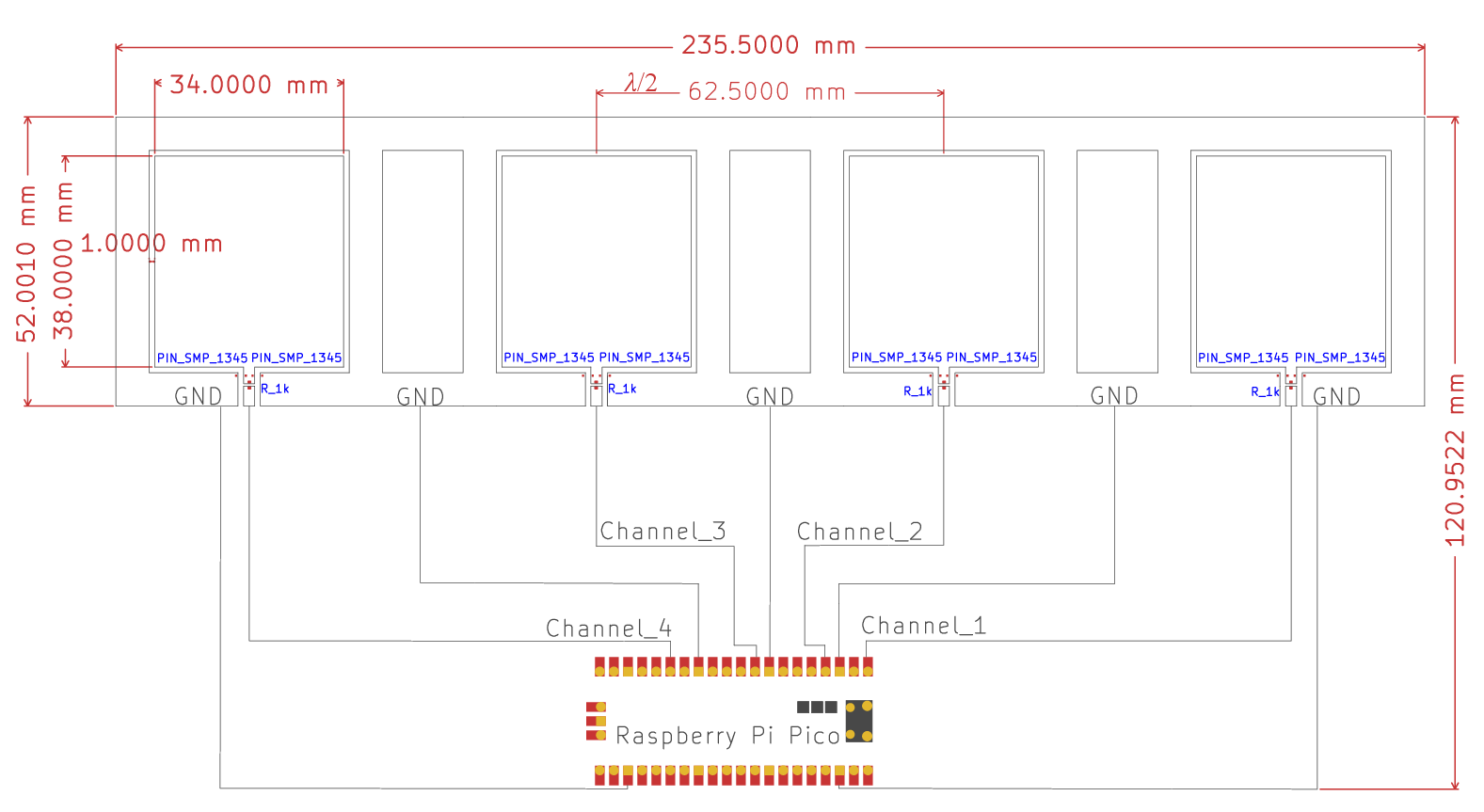}\label{fig:ris_schematic}}
    \hfill
    \caption{RIS. (a) Typical scenarios for sensing. (b) Prototype after additive manufacturing. (c) Schematic and footprint layout.}
    \label{fig:RIS}
\end{figure}
The design of the RIS, shown in Fig.~\ref{fig:ris_proto} and \ref{fig:ris_schematic}, comprises lumped elements mounted on a thin film inkjet-printed circuit board and coplanar waveguide (CPW) antennas~\cite{xie2024flexible}. 
At 2.45~GHz, the measured antenna return loss ($S_{11}$) is $-29.36$~dB with a 10~dB bandwidth of 575~MHz, and the measured maximum realized gain is 1.75~dBi with a half-power beamwidth 
of $96^{\circ}$.
The utilized thin film CPW antenna structure makes the antenna grounded in the same plane of the patch, without requiring an additional ground plane below the dielectric substrate. This allows the RIS elements to reflect and refract the incidence signals~\cite{zhang2022ios}. Such structure can also reduce the thickness and complexity of the RIS element.

Each RIS element contains a switching circuitry for load modulation. 
The switching circuitry utilizes a pair of PIN diodes SMP1345-079LF symmetrically connected to the patch and the ground of the antenna, switching the termination of the antenna between two loads $Z_{\rm L}^{(i)}\in\{Z_{\rm L}^{(1)},Z_{\rm L}^{(2)}\}$. 
In addition, a 1.0 k$\Omega$ surface-mounted device (SMD) resistor is applied between each baseband feeding and the diode pair for biasing the control signal.
The complex reflection coefficient~\cite{griffin2009linkbudget} is expressed by 
$\Gamma^{(i)} = (Z_{\rm L}^{(i)} - Z_{\rm a}^{*}) / (Z_{\rm L}^{(i)} + Z_{\rm a})$,
where $Z_{\rm a}$ is the antenna characteristic impedance of each RIS element, and $^{*}$ denotes the complex conjugation.
The load modulation results in two corresponding reflection coefficients, $\Gamma^{(1)}$ and $\Gamma^{(2)}$ over time. A $180^{\circ}$ phase difference between the two coefficients is expected for maximization of backscatter performance~\cite{griffin2009linkbudget}.
Simulated results of the reflection magnitude $\left|\Gamma^{(i)}\right|$ and phase $\angle\Gamma^{(i)}$ of the RIS element when the diode turns ON and OFF are shown in Fig.~2b using the equivalent circuit with measured antenna impedance and diode parameters~\cite{diode}. For instance, $Z_{\rm a}=46.85-j0.8\,\Omega$, $Z_{\rm L}^{(1)}=2.99+j4.02\,\Omega$, and $Z_{\rm L}^{(2)}=96.27-j508.72\,\Omega$ are adopted at 2.45~GHz.
It can be seen that $180^{\circ}\pm12^{\circ}$ phase difference (black curve) is achieved at the desired band from 2.4 to 2.5~GHz.
The reflection loss is less than 1~dB within the working frequency band.
The switch control signals are generated from an MCU Raspberry Pi RP2040, where the on-off keying (OOK) baseband modulation is implemented. 
\begin{figure}
\centering
\subfloat[]
{\includegraphics[width=0.48\columnwidth]{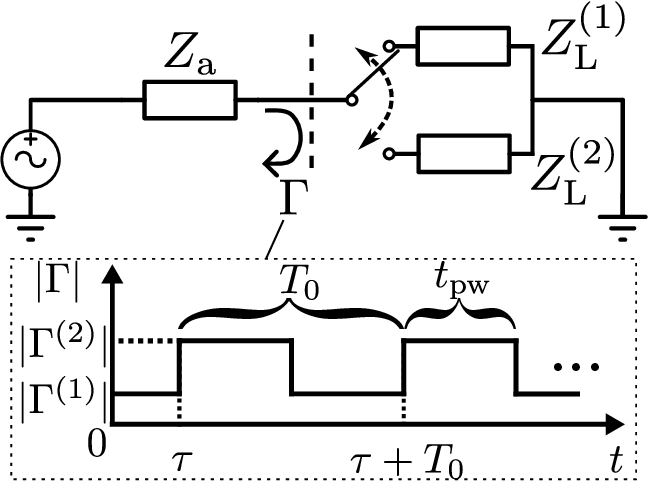}
    \label{fig:gamma_example}}
\hfill
\centering
\subfloat[]
{\includegraphics[width=0.48\columnwidth]{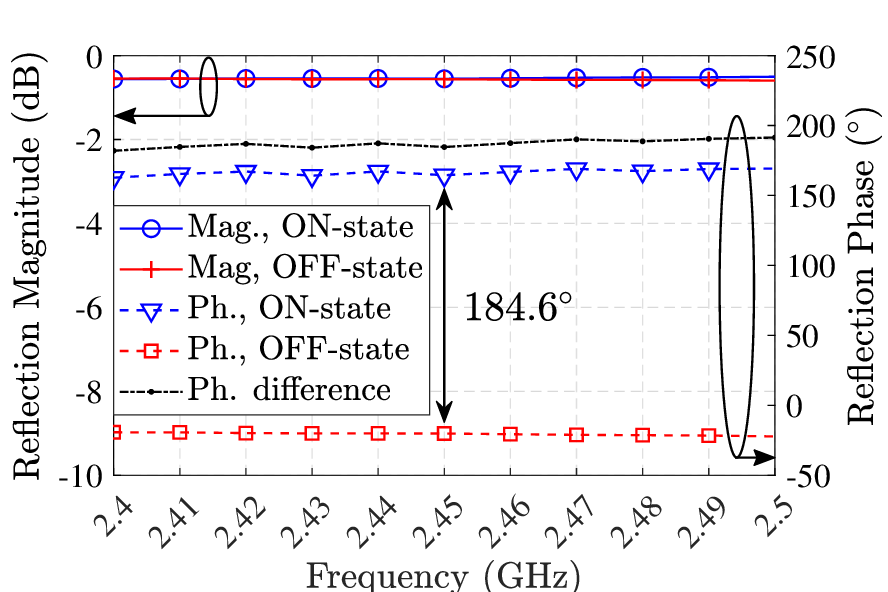}
    \label{fig:s11_compare_v3}}
\hfill
\centering
\subfloat[]
{\includegraphics[width=0.48\columnwidth]{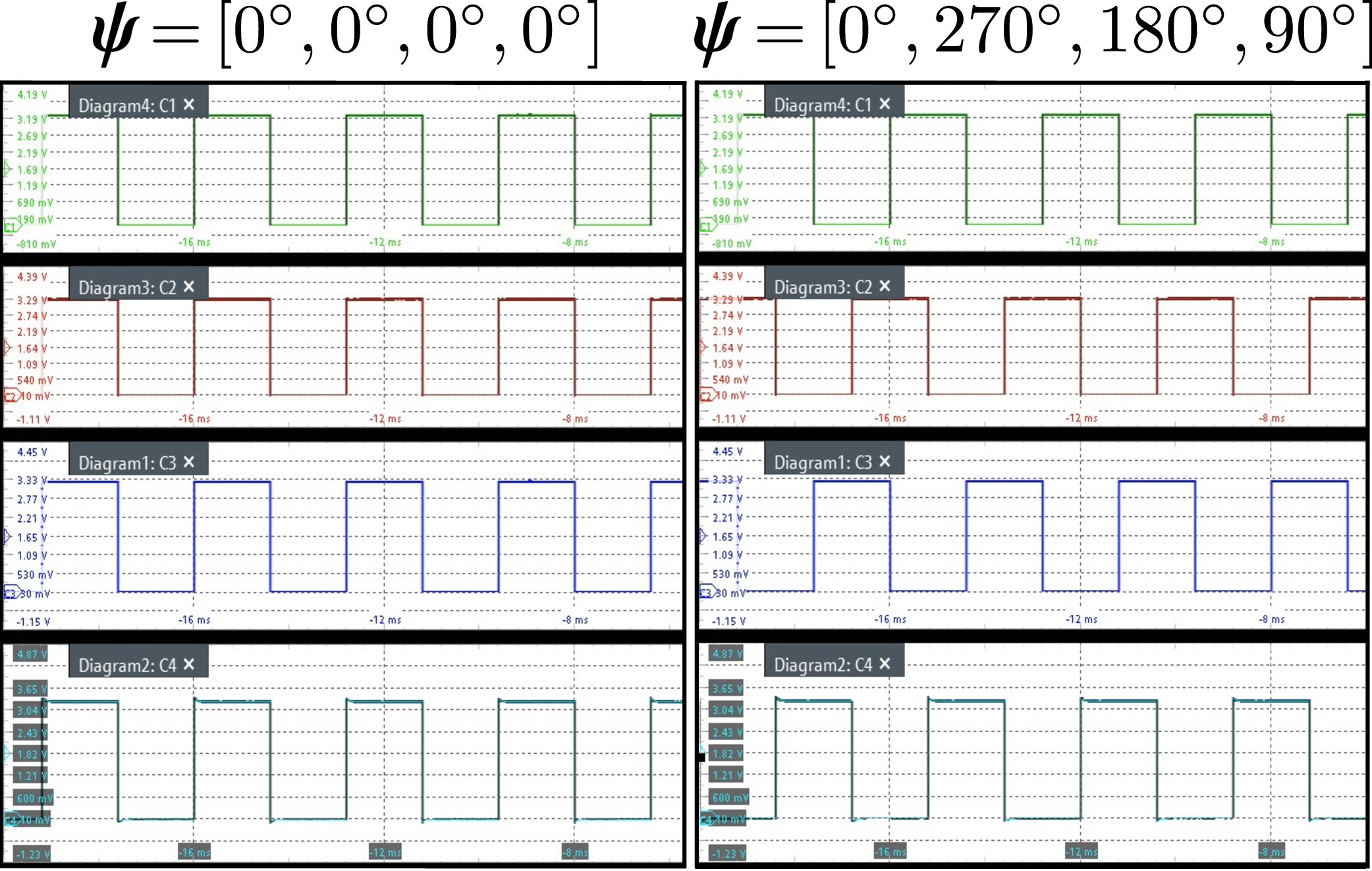}
    \label{fig:baseband_example}}
\hfill
\centering
\subfloat[]
{\includegraphics[width=0.48\columnwidth]{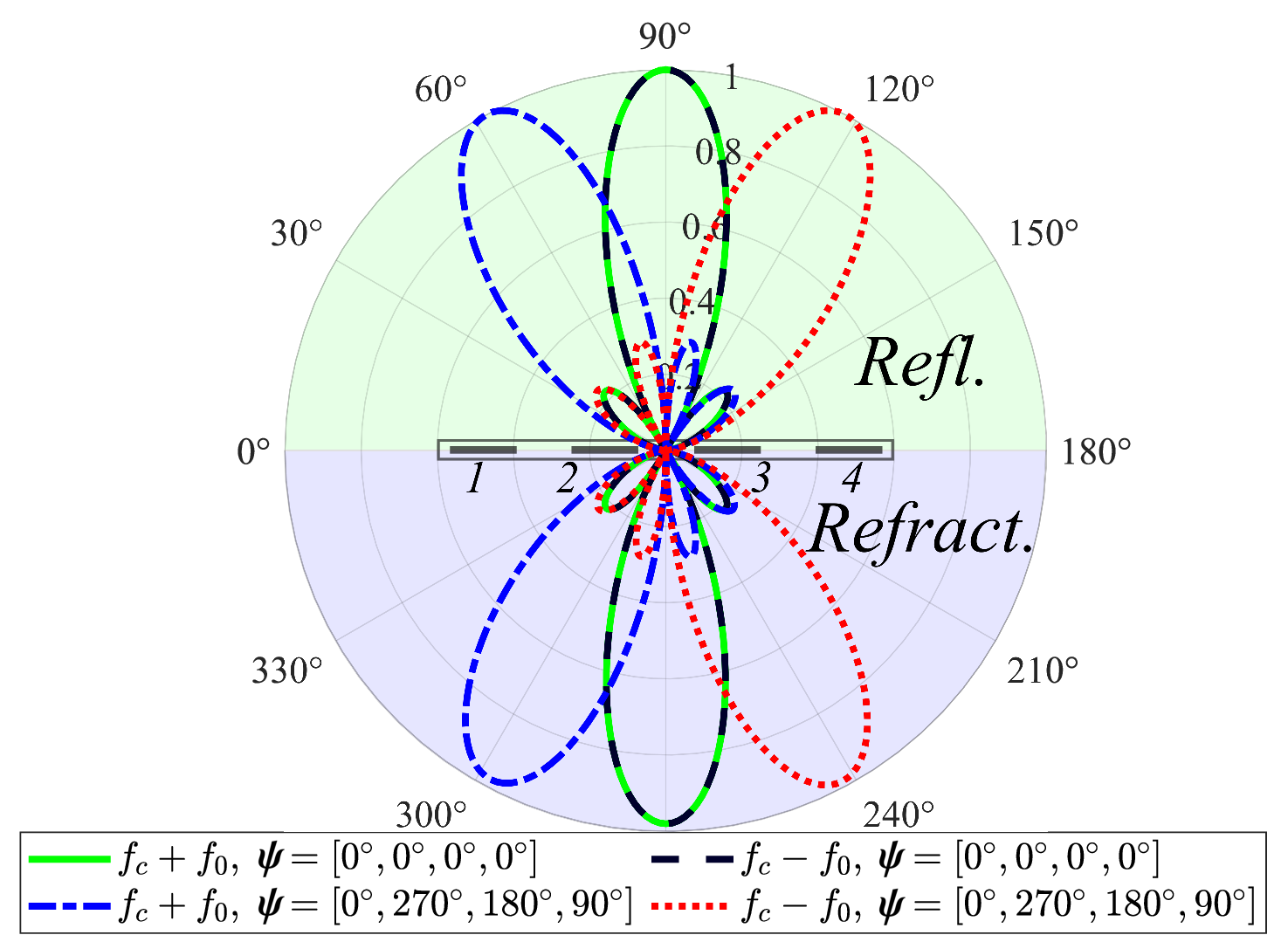}
    \label{fig:example_bf}}
\hfill
    \caption{Baseband-controlled beam steering. (a) Equivalent circuit of antenna load modulation. (b) Simulation of reflection magnitude and phase of the RIS element. (c) Phased baseband signals. (d) An example of normalized far-field RIS beam pattern with delay profiles.}
    \label{fig:baseband_control}
\end{figure}
The time-varying square-wave control signal results in the time-varying reflection coefficient $\Gamma(t)$. 
The Fourier series of $\Gamma(t)$, i.e., in frequency domain is expressed by
\begin{equation}
\label{eqn:1}
\Gamma(t)= \mathop {\sum}\limits_{m = - \infty }^\infty {c_{m}} e^{ {j2\pi mf_0t}},
\end{equation}
where $f_{0}$ is the frequency of control signal, and its Fourier series coefficients $c_{m}$ are
\begin{equation}
\label{eqn:2}
    \begin{split}
    c_{m}=\frac{1}{T_{0}}\int_{0}^{T_{0}}\Gamma\left(t\right)e^{-j 2\pi m f_{0} t}dt =
    \frac{1}{T_{0}} \Bigg( \Bigg. \int_{0}^{\tau}\Gamma^{(1)}e^{-j2\pi m f_{0} }{}^{t}dt \\ +\int_{\tau}^{\tau+t_{\rm pw}}\Gamma^{(2)}e^{-j2\pi m f_{0} }{}^{t}dt
    +\int_{\tau+t_{\rm pw}}^{T_0}\Gamma^{(1)}e^{-j2\pi m f_{0} }{}^{t}dt \Bigg. \Bigg) \\
    =\frac{j}{2\pi m}  (\Gamma^{(1)} - \Gamma^{(2)}) e^{-j2\pi m f_0 \tau } (1 - e^{-j2\pi m f_0 t_{\rm pw} }), 
    \end{split}
\end{equation}
where $T_{0}=1/f_{0}$ is the period of control signal, and $\tau\in[0, T_{0})$ is the time delay. In this paper, $t_{\rm pw}=T_{0}/2$ is the pulse width. Accordingly, the time delay $\tau$ brings a phase shift at the $m$-th ($m\neq0$) harmonic frequency. Fig.~\ref{fig:gamma_example} shows an example of the magnitude of time-varying $\Gamma$.
Under excitation from normal to the RIS surface direction by plane waves with a general expression $A_{\rm c}e^{j2\pi {f_{\rm c}}t}$ with an amplitude of $A_{\rm c}$ and frequency of $f_{\rm c}$, without loss of generality, the far-field scattering pattern of the RIS at the $m$-th harmonic frequency $f_{\rm c} \pm m f_{0}$ can be written as~\cite{zhang2018space}
\begin{equation}
\label{eqn:3}
    \begin{split}
    F_m\left( {\theta ,\varphi } \right) =& \mathop {\sum}\limits_{q = 1}^N \mathop {\sum}\limits_{p = 1}^M E^{p,q}\left( {\theta ,\varphi } \right)\\
    &\times e^{ {j\frac{{2\pi }}{{\lambda _{\mathrm{c}}}}\left[ {\left( {p - 1} \right)d_x{\mathrm{sin}}\;\theta\; {\mathrm{cos}}\;\varphi + \left( {q - 1} \right)d_y{\mathrm{sin}}\;\theta\; {\mathrm{sin}}\;\varphi } \right]}} \times c_{m}^{p,q}(\tau),
    \end{split}
\end{equation}
where $E^{p,q}\left( {\theta,\varphi } \right)$ is the far-field pattern of the $(p,q)$-th RIS element computed at the central frequency $f_{\rm c}$ and the corresponding wavelength $\lambda_{\rm c}$, which is approximated to be identical for all elements in this work.
$\theta$ and $\varphi$ are the elevation and azimuth angles, respectively, $d_{x}$ and $d_{y}$ are the element periods along the $x$ and $y$ directions, respectively.
For a desired pair of elevation and azimuth angle $(\theta, \varphi)$, 
by selecting an appropriate time-delay set $\bm{\tau} = \left[\tau_{1,1},\tau_{1,2},...,\tau_{1,N},...,\tau_{m,n},...,\tau_{M,N}\right] \in {\mathbb {C}^{MN}}$,
one can always find a maximum of $F_m$, resulting in a desired beam pattern
\begin{equation}
\label{eqn:4}
\begin{aligned}
    \underset{\bm{\tau}} {\text{maximize}} & & F_m (\theta, \varphi).
\end{aligned}
\end{equation}
By carefully selecting the frequency $f_{0}$ of the baseband control signals,
such method allows the RIS to steer scattered beams of the incidence RF signal using a limited bandwidth of $2f_{0}$, and hence the scattered signal might remain within the bandwidth of the incidence carrier signal. Furthermore, from the perspective of antenna theory, the radiation patterns can be approximated as identical at the central frequency $f_{\rm c}$ and its dominant harmonics $f_{\rm c}\pm f_{0}$ if $f_{0}$ is sufficiently small, such as in the range of hundreds of Hz.

In this work, we consider a $1\times4$ RIS model which is capable of steering the beam of scattered signal in azimuth directions. The MCU feeds four phased baseband control signals into each RIS element with delay profiles $\bm{\tau} = [\tau_{1},\tau_{2},\tau_{3},\tau_{4}]$ or their corresponding phases $\bm{\psi} = [\psi_1,\psi_2,\psi_3,\psi_4]$, respectively. The baseband signals have an approximately equal amplitude of 3.3~V and a frequency $f_{0}=313$~Hz.
The phase differences between elements can be configured from $0^{\circ}$ to $359^{\circ}$ with resolution of $1^{\circ}$.
Fig.~\ref{fig:baseband_example} shows an example of the MCU-generated 4 baseband signals monitored by an oscilloscope, where the left side shows an in-phase case when $\bm{\psi} = [0^{\circ},0^{\circ},0^{\circ},0^{\circ}]$ and the right side shows a case when $\bm{\psi} = [0^{\circ},270^{\circ},180^{\circ},90^{\circ}]$.
Calculated by~(\ref{eqn:3}), the normalized far-field scattering patterns corresponding to the above two delay profiles are shown in Fig.~\ref{fig:example_bf} with two dominant harmonics $f_{\rm c} \pm f_{0}$.
For $\bm{\psi} = [0^{\circ},270^{\circ},180^{\circ},90^{\circ}]$, the beam of $f_{\rm c}+f_0$ harmonic is steered to $60^{\circ}$ and $300^{\circ}$, whereas the beam of $f_{\rm c}-f_0$ harmonic is steered to $120^{\circ}$ and $240^{\circ}$, showing a symmetry with respect to $90^{\circ}$ and $270^{\circ}$. Such symmetry is valid for other delay profiles due to the nature of the symmetrical positive and negative components in~(\ref{eqn:1}).
For steering beams of the dominant harmonics $f_{\rm c} \pm f_{0}$ in different desired directions, their corresponding delay profiles of the baseband control signals are preloaded into the MCU.
\section{Manufacturing Process}
\begin{table}
\caption{Additive manufacturing procedure.}
\centering
\resizebox{\columnwidth}{!}{%
\begin{tabular}{ccccll}
\cline{1-4}
Procedure             & Materials                 & Equipment &Time consumption \\ \cline{1-4}
1. Inkjet printing    & SNP ink NBSIJ-MU0, PET substrate TP-3GU100                  & Inkjet printer Epson WF-2840 DWF & 20 sec\\
2. Photonic curing    & --          &  Xenon flash tube XOP-15, 1500 watt  & 15 sec \\
3. SMD mounting       & Epoxy adhesive MG Chemicals 9410      & Digital microscope and tweezers 
 & 15-20 min  \\
4. Oven curing        & --             &Oven Cecotec~610, 90 degree Celsius &60 min  \\
\cline{1-4}
\end{tabular}}
\label{tab:manufacturing}
\end{table}
A hybrid additive manufacturing method, which comprises inkjet printing with the use of commercial off-the-shelf components, is adopted to integrate the RIS circuitry and antennas on a flexible substrate. Such method has been implemented in~\cite{Boxuan-LED2024,wiklund2021review}, enabling cost-effectiveness and flexibility in design and production of compact and lightweight devices (here, W~236~$\times$~L~121~$\times$~H~1~mm and 9.9~g).
The manufacturing process includes (1) Inkjet printing of the circuit board and antenna; (2) Curing the ink on the printed sheet; (3) Mounting SMD components on the printed footprints with conductive adhesive, and (4) Curing the adhesive.
The utilized materials and equipment for the manufacturing procedure are listed in Table~\ref{tab:manufacturing}.
Silver nanoparticle (SNP) ink is used to print the conductive layer (thickness 440~nm, conductivity $6\times10^6$ S/m) of the design on an A4-size PET substrate (thickness 135~$\mu$m, relative permittivity 4.8, loss tangent 0.138) used with a low-cost inkjet printer. 
After inkjet printing, the printed surface undergoes a photonic curing to cure the ink. The photonic curing can mitigate thermal stress on the substrate. This reduction in thermal stress minimizes warping, degradation, and curing time, offering a distinct advantage over traditional convective oven curing methods that require high temperatures~\cite{kerminen2023cpw}. Following the photonic curing, SMD components are mounted to their footprints on the printed circuit board by using conductive silver-filled epoxy adhesive and thin tweezers.
Finally, for curing adhesive, the entire device is cured in a convection oven with $90^{\circ}$C for 60~minutes.
\section{Experiment and Evaluation}
\begin{table}
\caption{Numerical results of valid beam steering.}
\centering
\resizebox{\columnwidth}{!}{%
\begin{tabular}{ccccll}
\cline{1-4}
Desired dominance & Measured dominance & Measured dominance & \multirow{2}{*}{Delay profile $\bm{\psi}$} &      \\ 
direction at $f_{\rm c}+f_0$ & direction at $f_{\rm c}+f_0$ & direction at $f_{\rm c}-f_0$ & \\ \cline{1-4}
$50^\circ/310^\circ$       & $50^\circ/310^\circ$   & $140^\circ/220^\circ$  & $[0^\circ,244^\circ,129^\circ,13^\circ]$ &  \\
$60^\circ/300^\circ$       & $60^\circ/300^\circ$   & $130^\circ/230^\circ$  & $[0^\circ,270^\circ,180^\circ,90^\circ]$ &  \\
$70^\circ/290^\circ$       & $70^\circ/290^\circ$   & $110^\circ/250^\circ$, $120^\circ/240^\circ$  & $[0^\circ,298^\circ,237^\circ,175^\circ]$ & \\
$80^\circ/280^\circ$       & $80^\circ/280^\circ$   & $100^\circ/260^\circ$  & $[0^\circ,329^\circ,297^\circ,266^\circ]$ & \\
$90^\circ/270^\circ$       & $90^\circ/270^\circ$   & $90^\circ/270^\circ$   & $[0^\circ,0^\circ,0^\circ,0^\circ]$ &       \\
$100^\circ/260^\circ$      & $100^\circ/260^\circ$  & $80^\circ/280^\circ$   & $[0^\circ,31^\circ,63^\circ,94^\circ]$ &    \\
$110^\circ/250^\circ$      & $110^\circ/250^\circ$, $120^\circ/240^\circ$  & $70^\circ/290^\circ$   & $[0^\circ,62^\circ,123^\circ,185^\circ]$ &  \\
$120^\circ/240^\circ$      & $130^\circ/230^\circ$  & $60^\circ/300^\circ$   & $[0^\circ,90^\circ,180^\circ,270^\circ]$ &  \\
$130^\circ/230^\circ$      & $140^\circ/220^\circ$  & $50^\circ/310^\circ$   & $[0^\circ,116^\circ,231^\circ,347^\circ]$ & \\ \cline{1-4}
\end{tabular}}
\label{tab:setup}
\end{table}
\begin{figure}[t]
\centering
\includegraphics[width=0.98\columnwidth]{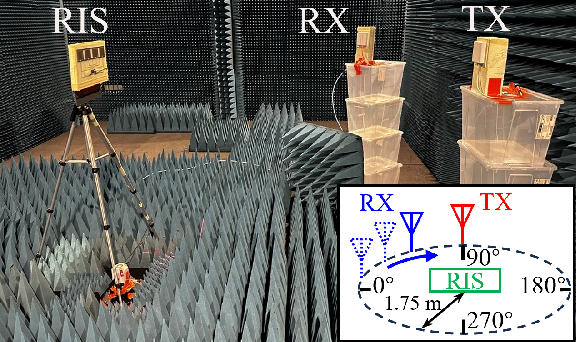}
\caption{Experimental setup.}
\label{fig_measurement_setup}
\end{figure}
\begin{figure*}
\centering
\null\hfill
\subfloat[]
{\includegraphics[width=0.98\columnwidth]{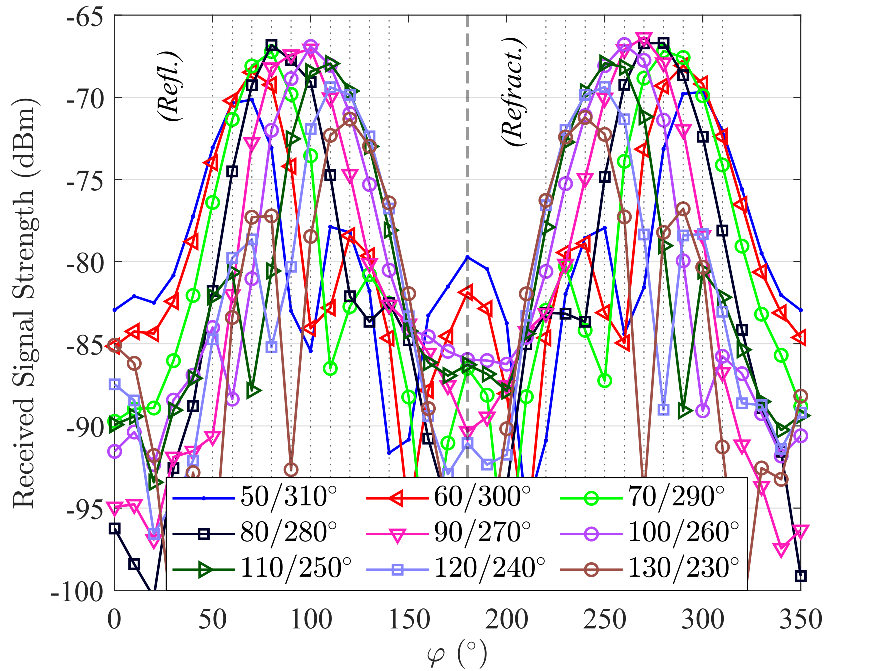}
    \label{fig:f1_pos_paper_wide_reduced}}
\hfill
\subfloat[]
{\includegraphics[width=0.98\columnwidth]{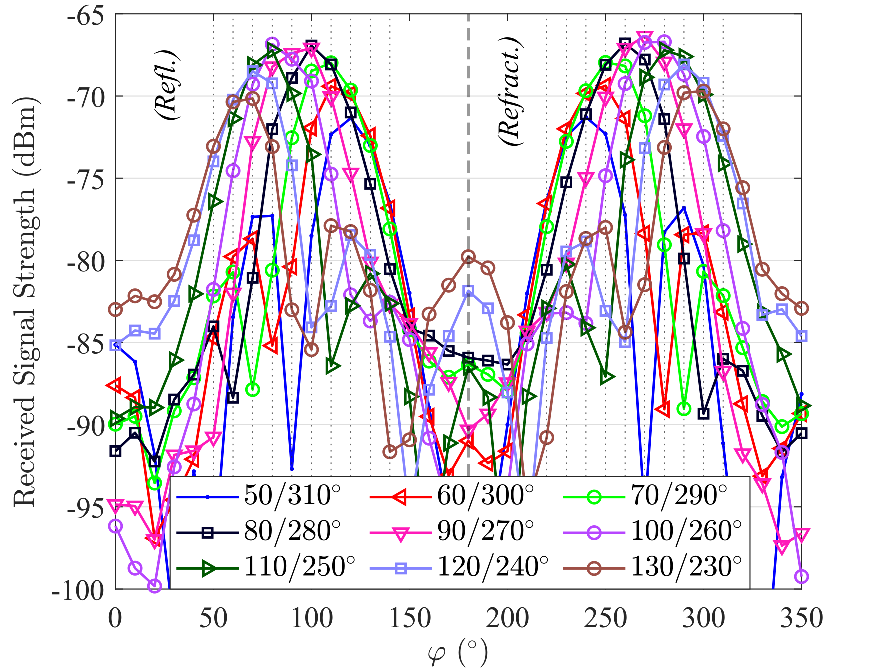}
    \label{fig:f1_neg_paper_wide_reduced}}
\hfill\null
    \caption{Harmonic beam steering measurement results. (a) +1st harmonic $f_{\rm c}+f_0$. (b) -1st harmonic $f_{\rm c}-f_0$.}
    \label{fig:results}
\end{figure*}
To validate the harmonic beam steering functionality of the RIS, the received signal strength of the harmonics was measured in a microwave anechoic chamber, with a measurement setup shown in Fig.~\ref{fig_measurement_setup}. 
A signal generator Rohde \& Schwarz SMBV100A was used as the carrier signal transmitter (TX) that transmits continuous wave at $f_{\rm c}=2.45$~GHz with power $P_{\rm c}=10$~dBm.
A real-time spectrum analyzer Siglent SSA3075XR was used as the receiver (RX) to detect harmonics in received signals.
Two reference patch antennas L-com HG2409P were used as measurement antennas of the TX and RX.
The RIS was located in the middle of the chamber and was connected to the MCU for dynamic control of the load modulation that occurred on the RIS elements.
Both measurement antennas were situated at a distance $r=1.75$~m from the RIS position.
During the measurement process, the position of the RX antenna was changed accordingly to the measured angle, while the location of the TX antenna was statically installed at $\varphi=90^\circ$ to form normal plane wave excitation.
Scattered signals were sampled with $\Delta\varphi=10^\circ$ intervals. 
For each RX position and delay profile, the received signal strength of $\pm1$st harmonics was measured.
Table~\ref{tab:setup} lists all the measured cases corresponding to delay profiles. 
Delay profiles for nine reflection/refraction angles of +1st harmonic of the scattered signal were formed based on~\eqref{eqn:4}.
The angles were chosen based on the beamwidth of the single patch element of the structure under illumination.
The presented $1\times4$ RIS has a diversity gain of around 9 dB from the broadside radiation, compared with the single-element model~\cite{xie2024flexible}.
The measured average power consumption of the RIS controller MCU RP2040 is 122~mW.

Fig.~\ref{fig:results} presents the measured received signal strength of $\pm1$st harmonics with frequencies of $f_{\rm c}\pm f_0$. 
It is visible that the manufactured RIS fulfills the proposed function, where the beam steering for both harmonics depending on the phase profiles is achieved.
Both reflection (front-side, $50^{\circ}$-$130^{\circ}$) and refraction (back-side, $230^{\circ}$-$310^{\circ}$) phenomenon of the harmonics are observed, showing a symmetry with respect to the RIS plane ($0^{\circ}$/$180^{\circ}$).
Table~\ref{tab:setup} tabulates the angles at which each of the measured profiles provided the maximal received power for both harmonics. 
As shown in Fig.~\ref{fig:f1_pos_paper_wide_reduced}, for the majority of profiles, +1st harmonic has the highest value among the measured profiles in the intended direction. 
A similar phenomenon of -1st harmonic can be observed in Fig.~\ref{fig:f1_neg_paper_wide_reduced}.
Furthermore, comparing the +1st and -1st harmonic, the symmetry is observed with respect to $90^\circ/270^\circ$, which validates the nature of the symmetrical positive and negative components proposed in~(\ref{eqn:1}).
The results demonstrate the possibility of controlling the beamforming process following the proposed design specifications, which is useful for the development of such kind of low-complexity RIS.
A deterioration of the received power with an increase of incline from the normal directions can be explained by the decrease of the RIS single element gain. 
However, some delay profiles, e.g., with the desired direction of scattering $120^\circ/240^\circ$ did not exhibit the optimal performance among measured profiles in the intended angle. 
This is attributed to the assumption of identical RIS element patterns, introduced to simplify~\eqref{eqn:3} during optimization~\eqref{eqn:4}.
In fact, due to the proximity of the ports and the sharing of common structural elements, the coupling between four individual elements of RIS is not negligible. As a result, the radiation patterns of elements are not perfectly equal and hence affect the optimization strategy in~\eqref{eqn:4}. 
To address this issue, the RIS element design can be further optimized, considering the coupling and adjacent interactions between elements. Moreover, adaptive calibration algorithms can be implemented which dynamically adjust to observed discrepancies between expected and actual beam steering patterns.

The harmonic beam steering technique can enhance RIS-aided sensing applications, such as indoor localization and assisted living~\cite{zhang2022sensing}, by mitigating RF interference at the carrier frequency and improving spectrum utilization.
By employing the harmonics, this technique simplifies the RIS design, preserves the integrity of carrier communication channels in dense spectral environments, and efficiently utilizes underexploited spectral regions for sensing purposes. 
Furthermore, such method can control beams of both the positive and negative harmonics simultaneously, and hence sense separated users at the same time.
These advancements position harmonic beam steering as a superior alternative to conventional beamforming methods.
Future research on the RIS can explore several key areas to enhance its functionality. 
Optimizing the element structure and involving a suitable superstrate can improve beamforming gain through radar cross-section reduction~\cite{costa2014rcs}.
Furthermore, employing adaptive calibration techniques can dynamically adjust the beamforming parameters to optimize beam patterns in real-time~\cite{pei2021adaptive,he2020adaptive}.
Moreover, expanding RIS to include hundreds or even thousands of elements would allow for more precise signal manipulation and broader beam-steering capabilities.
As the scale of RIS increases, resource allocation strategies such as partitioning into sub-surfaces~\cite{kumar2024partition} will be essential for serving and sensing multiple users efficiently.
In addition, investigating spectral and energy efficiency in RIS-enabled integrated sensing and communication (ISAC) systems is also critical for future energy-efficient networks.
\section{Conclusion}
This letter has presented a thin film RIS, leveraging inkjet printing and additive manufacturing on a flexible substrate to offer a low-cost and low-complexity RIS solution. 
The RIS design has integrated antennas, switching circuitry, and an MCU, without a ground shield.
The configuration enables individual and simultaneous control of RIS elements, manipulating the captured carrier signal by reflecting and refracting its dominant harmonics.
The proposed harmonic beam steering functionality has been validated that can dynamically steer the beams toward multiple desired directions at both front and back sides of the RIS.
Results show that the proposed device has the potential to enable RIS-aided communication and sensing applications.
\normalsize
\bibliographystyle{IEEEtran}
\bibliography{reference}
\end{document}